\documentstyle[aps,prl,multicol,epsf]{revtex}

\begin{document}

\title{Viscoelastic Depinning of Driven Systems: Mean-Field Plastic Scallops}

\author{M. Cristina Marchetti, A. Alan Middleton, and Thomas Prellberg}
\address{Physics Department, Syracuse University, Syracuse, NY 13244}
\date{December 23, 1999}

\maketitle

\begin{abstract}
We have investigated the mean field dynamics of an overdamped
viscoelastic medium driven through quenched disorder.
The model allows for the coexistence of pinned and sliding
regions and can exhibit continuous elastic depinning or first order
hysteretic depinning.
Numerical simulations indicate mean field instabilities that
correspond to macroscopic stick-slip events and lead to premature switching.
The model describes the elastic and plastic dynamics of driven vortex
arrays in superconductors and other extended disordered systems.
\end{abstract}
\begin{multicols}{2}

Extended condensed matter systems driven over quenched disorder 
exhibit a very complex dynamics, including
nonequilibrium phase transitions and history dependence.
Such systems include vortex arrays in type-II
superconductors\cite{blatter94}, charge density waves in anisotropic
conductors\cite{gruner88,fisher85}, and many others.
Closely related behavior also arises in friction
and lubrication\cite{persson}, where a surface or monolayer is brought
in contact with another solid surface and forced to slide relative to
it. 

Most of the theoretical work to date has focused on the
dissipative dynamics of driven
{\it elastic} media that are distorted by disorder, but cannot
tear. At zero temperature such systems exhibit a sharp depinning
transition from a pinned 
to a sliding
state\cite{fisher85,AAMthesis}. The transition, first studied in
the context of charge density waves, 
is continuous, with universal
critical behavior. The sliding state is unique and there
is no hysteresis or history dependence \cite{nocross}.  More
recent work, while still focusing on elastic media, has shown that the
dynamics is quite rich well into the uniformly sliding
state\cite{fisher97,giamarchi96,bmr98,moon96,pardo98}. 

On the other hand, experiments\cite{shobo93,henderson98} and simulations
\cite{shi91,faleski96} show that the elastic medium model is
inadequate for many physical systems with strong disorder that upon
depinning exhibit a spatially inhomogeneous plastic response, without
long wavelength elastic restoring forces. In this plastic flow regime,
topological defects proliferate and the system is broken up in
fluid-like regions flowing around pinned solid regions.  
Not much progress has been made in describing this behavior
analytically.  The wealth of experimental work on driven vortex arrays
clearly indicates that, in most of the field and temperature region of
interest, the current-driven vortex dynamics 
is strongly history dependent, with
long-term memory and switching as the system explores a variety of
nonequilibrium sliding states \cite{shobo93,henderson98}.

In this paper we describe a {\it coarse-grained} model for the
dynamics of a driven medium that allows for spatially inhomogeneous
response, with the coexistence of moving and pinned regions.
The model is inspired by the well-known phenomenology
of viscoelasticity in dense fluids \cite{boon92}.
The elastic couplings between the local displacements are
replaced by couplings that are nonlocal
in time and allow for elastic restoring forces to turn into
dissipative fluid flow on short time scales. The model yields
elastic depinning in one limit; as the parameters are varied, it
incorporates continuous depinning, hysteretic plastic depinning and eventually viscous flow, allowing
the crossovers between these regimes (such as those,
observed in vortex arrays \cite{shobo93,faleski96})
to be studied in detail.
For
a wide range of parameter values the depinning transition
is first order, with
velocity hysteresis (switching.)  The nonlinear velocity-force
characteristic can be evaluated analytically in mean field for various types of
pinning forces, under the assumption of constant mean field velocity.
Numerical simulations confirm
the inhomogeneous nature of the dynamics, with pinning and tearing
(coexisting moving and pinned degrees of freedom.)
In addition, the mean velocity
near depinning fluctuates, due to {\em macroscopic}
stick-slip type events. These events appear to only mildly
violate the uniform mean-velocity assumption but directly lead to
switching from one velocity branch to another before the first branch
terminates (premature switching.)
Models that account for switching in charge
density waves and are in spirit similar to ours have been proposed and
studied by Strogatz and
collaborators \cite{strogatz}.  In such models, plasticity is modeled
by a non-convex elastic potential, in contrast with the velocity
convolutions studied here. A model similar to ours has also been proposed
for crack propagation \cite{schwarzfisher}.

{\it The model: a driven viscoelastic medium.}  To motivate our model,
we first recall the generic model of driven {\it elastic}
media \cite{fisher85} discussed extensively in the literature,
where the long-wavelength dynamics is
described in terms of a coarse-grained displacement field, $u({\bf r},t)$.
The displacement fields  represent deformations of regions pinned 
collectively by disorder
(e.g., a Larkin domains) and
are coupled by convex elastic interactions. No
topological defects are allowed.
Considering, for simplicity,  the overdamped dynamics of a scalar field (the model
is easily extended to the more general case) 
and modeling the displacement field on
lattice sites,
$u({\bf r},t)\rightarrow u_i(t)$,
the equation of motion for the
local field
$u_i$ (measured in the laboratory frame\cite{note_conv}) at site $i$ 
is 
\begin{equation}
\label{elastic}
\gamma_0\dot{u}_i=\sum_{\langle
ij\rangle}\mu_{ij}~(u_j-u_i)+F+F_i(u_i),
\end{equation}
where the summation is restricted to nearest neighbor pairs and $\gamma_0$ 
is the friction.
If all the nearest-neighbor elastic
couplings, $\mu_{ij}\geq 0$,  are equal, the first term on the right hand
side of Eq.\ (\ref{elastic}) is the discrete Laplacian in $d$
dimensions.  The second term is the homogeneous driving force, $F$,
and $F_i(u_i)$ denotes the pinning force arising from a quenched
random potential, $V_i(u_i)$,
%
$F_i(u_i)=-{dV_i / du_i}=h_i f(u_i-\beta_i)$,
%
with $f(u)$ a periodic function with period $1$ and $\beta_i$
random phases uniformly distributed in $[0,1]$.
The $h_i$ are chosen independently at each site
from a distribution $\rho(h)$.
One of the quantities of interest is the average velocity of
the driven medium,
$\overline{v}(F)=N^{-1}\sum \dot{u}_i$.
For an elastic medium there is a unique stationary sliding state for
$F>F_c$, with critical behavior $\overline{v}(F)\sim (F-F_c)^\beta$ \cite{fisher85},
and no hysteresis at the transition \cite{nocross}.

We now modify the elastic interactions in Eq.\ (\ref{elastic}) to allow
for local tearing of the medium. Inspired by standard models of
viscoelasticity, we replace the elastic interaction
by  couplings to the local velocity field,
$v_i=\dot{u}_i$, that are nonlocal in time.  Our model equation for the overdamped dynamics of
a driven viscoelastic medium is \cite{footbeta}
\begin{equation}
\label{viscoel}
\gamma_0\dot{u}_i=\sum_{\langle ij\rangle}\int_0^t
ds\,C_{ij}(t-s)\big[\dot{u}_j(s)-\dot{u}_i(s)\big]
+F+F_i(u_i),
\end{equation}
where the viscous couplings $C_{ij}(s)$ have finite first
moments,
$\int_0^\infty ds\,C_{ij}(s)=\eta_{ij}<\infty$
and $C_{ij}(0)=\mu_{ij}$.  
Such nonlocal couplings to velocity  are of course not present at
the microscopic level, but are generated generically upon coarse-graining\cite{boon92,mcmvisco}.
Eq.\ (\ref{viscoel}) is a coarse-grained model for the dynamics
of a driven disordered medium that allows for slip or friction 
of the interacting Larkin domains relative to each other.

A simple, yet successful, model of viscoelasticity due to Maxwell
is obtained when the memory kernels are assumed to be uniform in
space and to decay exponentially in time, according to
$C_{ij}(t)=\mu e^{-t/\tau}$,
with $\tau=\eta/\mu$ the Maxwell relaxation time.  
For $\tau\rightarrow\infty$ and fixed $\mu$, Eq.\ (\ref{viscoel})
reduces to Eq.\ (\ref{elastic}) for a driven elastic medium.  For
$\tau\rightarrow 0$ and $\eta$ fixed,
the first term on the right hand side of
Eq.\ (\ref{viscoel}) can be approximated as $\eta\sum_{\langle
ij\rangle}[v_j(t)-v_i(t)]$, which represents viscous forces coupling
the local fluid velocity at different spatial points.  In this limit,
Eq.\ (\ref{viscoel}) describes an overdamped lattice-fluid of viscosity
$\eta$. We propose
Eq.\ (\ref{viscoel}) as a simple, yet realistic model for a
driven disordered system that exhibits spatially inhomogeneous plastic
response.

{\it Mean Field Approximation.}  As for the driven elastic
media, substantial
analytical progress in two or three dimensions is presumably
only possible via perturbation theory or
by a functional renormalization group treatment \cite{percnote}.
An alternative
approach that has provided valuable insight 
for a
driven elastic medium is mean field theory (MFT), first discussed
by D.\ S.\ Fisher \cite{fisher85}.
MFT is formally valid in the limit of infinite-range interaction, with
$\sum_j C_{ij} = N C(t)$ held fixed.
The equation of motion for the displacement at each site is then
given by
\begin{equation}
\label{viscoel_MF}
\gamma_0\dot{u}_i=\int_0^t
yds\,C(t-s)\big[\overline{v}(s)-\dot{u}_i(s)\big]+F+F_i(u_i),
\end{equation}
where the mean field is given by
$\overline{u}(t)= N^{-1}\sum_{i=1}^N u_i(t)$,
and
$\overline{v}(t)=\dot{\overline{u}}(t)$.

If the memory kernel $C(t)$ is chosen to be of the Maxwell form,
it is then possible to transform the
integro-differential equation (\ref{viscoel_MF}) to a second-order
differential equation, given by
\begin{eqnarray}
\label{final_eqn}
\tau\ddot{u}+\gamma(\eta,\tau,h;u_i)\dot{u}=F+F_i(u_i)+\eta\overline{v},
\end{eqnarray}
with
$\gamma(\eta,\tau,u_i;h)=1+\eta - \tau{\partial F_i\over\partial
u_i}$
an effective friction.
We have scaled  Eq.\ (\ref{final_eqn}) 
by letting 
$\tau\rightarrow\tau h_0$, $t\rightarrow th_0$, $\eta\rightarrow \eta/\gamma_0$,
$F\rightarrow F/(\gamma_0 h_0)$ and $h\rightarrow h/(\gamma_0 h_0)$,
where
$h_0$ is the characteristic scale of the distribution $\rho(h)$.
With this change of variables, 
the model is now characterized by two parameters, $\eta$ and $\tau$, and the shape
of $\rho(h)$.
The MF equation for
our viscoelastic model closely resembles the MF equation for
a driven {\it massive} elastic medium, with $\tau$ playing the role of
the mass. The most important difference is that in the massive elastic
medium the MF term $\eta\overline{v}$ is replaced by
$\mu\overline{u}$. As a result, the MFT of a driven massive elastic
medium even with constant $\overline{v}$
contains three degrees of freedom (as opposed to the two of our
problem) and the dynamics of a single $u_i$ is chaotic
\cite{braun,strunz}.

We are first interested here in steadily sliding
solutions of the MF model,
Eq.\ (\ref{viscoel_MF}).  It is
natural to look for periodic solutions $u_p(t;h)$ of period $T(h)$,
($\int_0^{T(h)}\,dt\,\dot{u}_p(t;h)=1$) that may set in after
an initial transient ($t\gg T(h)$).  Such solutions need not
be unique.
Guided by a large body of previous work on driven elastic media, we
focus on the MFT for the case of
a pinning potential with cusp-like singularities, which better captures
the physics of the corresponding
finite-dimensional model \cite{MFnote}.
An explicit solution of Eq.\ (\ref{final_eqn}) can be obtained for the
scalloped parabolic potential,
$V(u)=(h/2)(u^2-u+1/4)$.
In this case Eq.\ (\ref{final_eqn}) is linear
and its general solution is
%
$u_p(t;h)=C_1\exp(-\lambda_1t)+C_2\exp(-\lambda_2t)+ {1 / 2}+(\eta
\overline{v}+F)/ h$,
%
with
%
$\lambda_{1,2}=\left(1+\eta+\tau h\pm\sqrt{(1+\eta+\tau h)^2-4\tau h}\right)/(2\tau)$.
%
For each fixed value of $h$, we obtain an implicit equation for the
period, $T(h)$,
%
$\eta\overline{v}+F={\cal G}(T;\eta,\tau,h)$,
%
with
\end{multicols}
\begin{equation}
\label{bigG}
{\cal
G}(T;\eta,\tau,h)={\lambda_1(1-e^{-\lambda_1T})-\lambda_2(1-e^{-\lambda_2T})
+\tau\lambda_1\lambda_2(e^{-\lambda_1T}-e^{-\lambda_2T})\over
(\lambda_1-\lambda_2)(1-e^{-\lambda_1T})(1-e^{-\lambda_2T})} -{h\over
2}.
\end{equation}
\begin{multicols}{2}
\noindent The solution of Eq.\ (\ref{bigG}), together with
the self-consistency
constraint $\overline{v}=<[T(h)]^{-1}>_h$, determines the drift velocity as a function of
driving force, $F$.  When $T(h_i)\rightarrow\infty$, the $u_i$ is pinned.

Figure 1 shows the analytical solution for the mean velocity 
as a function of driving force for $\rho(h)=\exp(-h)$. 
The depinning occurs at $F=0$ for all distributions of pinning strengths, 
$\rho(h)$, with support not bounded from below by
a positive $h_{\rm min}$. 
For small $\eta$ and $\tau$,
corresponding to weak coupling among the local displacements, 
the analytical solution is single-valued
and the depinning is continuous. 
For large $\eta$ and $\tau$ the analytical solution yields 
multi-valued velocity curves,
reflecting the existence of multiple sliding states,
and the depinning is hysteretic.
As shown in the inset of Fig.~1, there is a critical value, $\eta_c(\tau)$, 
that separates single-valued from multi-valued solutions.
The value $\eta=\eta_c$ is a critical point and the velocity curve is expected to exhibit critical
scaling. While the value of $\eta_c$ depends on $\tau$, the existence of an hysteretic region
at large $\eta$, with coexistence
of sliding and moving states and
early switching (see also Fig.\ 1)
occurs for all finite values of $\tau$, including $\tau=0$.
For
$\tau\rightarrow\infty$ and $\eta\rightarrow\infty$, with the ratio 
$\mu=\eta/\tau$ held fixed,
Eq.\ (\ref{final_eqn}) reduces to the MFT of an overdamped elastic
medium \cite{footelastic}.  In this case an analytical
solution is available 
and the velocity vanishes linearly as $F\rightarrow F_c$  \cite{NF92}.

%
%

\begin{figure}
\begin{center}
\leavevmode
\epsfxsize=7.2cm
\epsffile{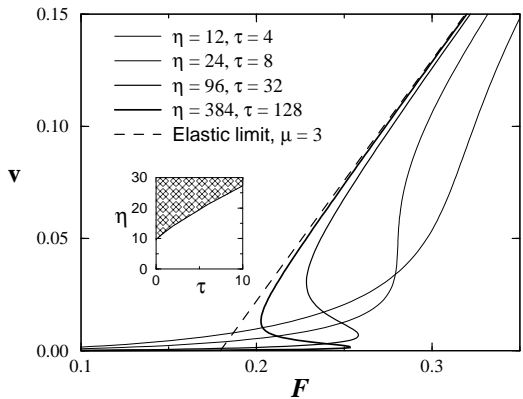}
\end{center}
\caption
{Analytical MFT velocity versus
force curve for $\rho(h)=\exp(-h)$
and selected values of $\tau$ and $\eta$ along the path $\eta=3\tau$
in parameter space. As $\tau$ and $\eta$ increase along this path,
the $v(F)$ curve becomes hysteretic.
The dashed line is the result for the purely
elastic model with $\mu = 3$, showing the convergence of the viscoelastic
model to the elastic model for large $\tau$ and $\eta$.
The inset shows the regions of parameters where depinning is
hysteretic (shaded region) and where it is continuous (unshaded region.)
}
\label{figure1}
\end{figure}

{\it Numerical work.} We have investigated the stability of the
branches of the analytically determined current-drive relationship.
We performed direct numerical simulation of the equations of motion,
for both force drive and constrained mean velocity.
The simulations
were performed using two codes, for verification:
a Runge-Kutta integration and an event-driven Euler integration,
with the ``events'' being crossings of a displacement $u_i$
from one parabolic region to the next.
The results were checked for insensitivity to time step $\Delta t$ and
size $N$. For the constant $\overline{v}$ constraint, the drive-velocity
relationship matches the analytical prediction.

\begin{figure}
\begin{center}
\leavevmode
\epsfxsize=7.2cm
\epsffile{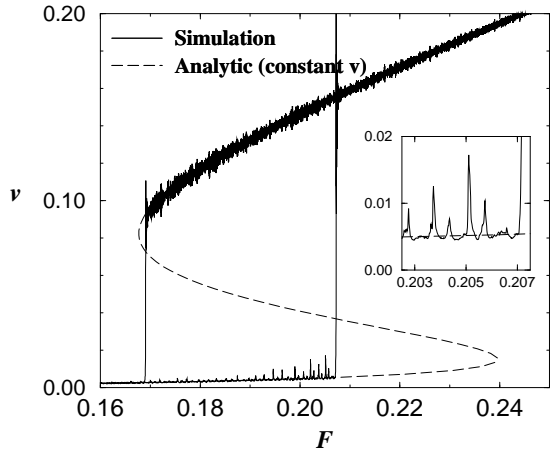}
\end{center}
\caption{Comparison of direct numerical simulation (solid line)
with analytic predictions (dashed line),
which assume a constant $\overline{v}$, for $\rho(h)=\exp(-h)$,
$\eta = 32$, $\tau=0.8$,
$N=16384$, and a ramp rate of $dF/dt = 2.5 \times 10^{-6}$. The field $F$
is cyclic in time. The results are
in near exact agreement for much of the history. In the hysteretic region,
on the lower branch,
the mean field velocity occasionally spikes due to macroscopic
events. 
}
\label{figsimul1}
\end{figure}

In the regions where the velocity is a unique function of the drive,
the simulation results with slowly changing $F$
for the force-drive curve match very closely those
of the analytical results, which assume a constant $\overline{v}$.
In the presumed hysteretic region, though, the simulation results can be
quite different. In particular, we note two features: mean field velocity
oscillations on the lower branch and ``early'' switching, where the
mean velocity switches from the lower to upper branch prior to the end
of the analytically computed hysteresis region.
A sample hysteresis curve indicating early switching is shown in
Fig.\ \ref{figsimul1}. We have computed the
magnitude of the fluctuations in the mean
velocity on the upper branch as a function of $N$: the
results are numerically consistent with a magnitude $\propto N^{-1/2}$,
indicating that these fluctuations vanish as $N\rightarrow\infty$.
The fluctuations on the lower branch do {\em not} vanish in the limit
of large $N$, however. These fluctuations are presumably due to an
instability of the constant $\overline{v}$ solution in the large volume
limit.
We hypothesize, with the support of detailed analysis of the
numerics, that {\it nearly depinned}
degrees of freedom
(which would remain pinned at constant $\overline{v}$) are
made unstable by velocity fluctuations and lead to an
avalanche type of behavior, which causes a peak in $\overline{v}$.
The magnitude of this instability apparently becomes large enough to
drive the mean velocity to the upper branch before the
presumed constant $\overline{v}$ velocity jump occurs.

In conclusion, we have introduced a coarse-grained model of plastic
flow that allows for slip of coherently pinned domains. We have solved this
model analytically in mean field for the case of Maxwellian kernel, under the
assumption of non-fluctuating mean velocity. 
We find that (1) the model exhibits both continuous and first order hysteretic
depinning as the parameters are varied, (2) we can recover the case of elastic
depinning in one limit, (3) pinned and sliding regions coexist in the hysteretic regime, 
and (4) the mean velocity curves display features observed in experiments.
Numerical simulations suggest
that the behavior is much richer than suggested by the MF calculation and
includes stick-slip-like instabilities which lead to early switching.
Strong history dependence has been observed in the dc response 
of vortex lattices in type-II superconductors \cite{shobo93,henderson98}
and in charge density waves \cite{Lemay}.
Hysteresis in vortex lattice motion
is most pronounced in the region of the so-called 
peak effect, where the dc response during
ramp-up of the current proceeds
via a series of jumps. These have been attributed to strong spatial
inhomogeneities in the distribution of vortex velocities, not
unlike what is observed in our model \cite{conserve}. 
We expect that in finite dimensions, the transition to hysteresis
will be characterized by non-trivial universal scaling exponents \cite{inprepMM},
similar to the situation for hysteresis in random magnets \cite{percnote},
and that these exponents could be experimentally tested.

One of us (MCM) thanks Daniel Fisher and Jennifer
Schwarz for illuminating discussions.
The work was supported by NSF through grants DMR-9730678,
POWRE-DMR9805818 and CAREER-DMR9702242.

\end{multicols}


\begin{references}
\bibitem{note_conv}
We neglect here convective derivatives, which may indeed be
important near a first order transition.
\bibitem{footbeta}
The disorder $\beta_i$ can be absorbed into the initial
conditions for $u_i$,
as the coupling is only through $\dot{u}_i$.
Non-uniform initial conditions should
then be chosen to reflect this disorder,
otherwise degenerate solutions arise.
\bibitem{MFnote}
Numerical simulations \cite{AAMthesis} and functional RG treatments \cite{NF92}
have shown that in low dimensions 
the critical region exhibits
a jump-dominated jerky dynamics 
even when the pinning potential
is smooth on short scales.
In MFT, where the force is averaged over many neighbors, this
crucial feature of the dynamics survives only for pinning potentials with cusp-like
singularities.
\bibitem{conserve} In finite dimensions, a conservation law would be
added to most properly describe vortex motion, while this is not necessary
for the charge density waves condensate.
\end{references}
\end{document}